\documentclass[10pt]{asme2ej}
\usepackage{epsfig} 
\usepackage{amsmath}
\usepackage{accents}
\usepackage{amssymb}
\usepackage{physics}
\usepackage{commath}
\usepackage{tikz}
\usepackage{url}
\usepackage{soul}
\usepackage{multirow}
\newcommand\ds{\displaystyle}
\graphicspath{{svg/}}
\def\dyadic{\otimes}

\title{Anisotropic multicomponent topology optimization for additive manufacturing with build orientation design and stress-constrained interfaces\footnote{Preliminary version of this manuscript has been presented in ASME 2019 Computers and Information in Engineering Conference, August 18 – 21, Anaheim, CA.}}

\author{Yuqing Zhou
	\affiliation{
		University of Michigan\\
		Ann Arbor, Michigan 48109\\
		Toyota Research Institute of North America\\
		Ann Arbor, Michigan, 48105\\
		Email: yuqingz@umich.edu
	}
}
\author{Tsuyoshi Nomura
	\affiliation{
		Toyota Research Institute of North America\\
		Ann Arbor, Michigan, 48105\\
		Toyota Central R \& D Labs., Inc.\\
		Yokomichi, Nagakute 480-1192, Japan\\
		Email: tsuyoshi.nomura@toyota.com
	}
}
\author{Kazuhiro Saitou
	\affiliation{
		University of Michigan\\
		Ann Arbor, Michigan 48109\\
		Email: kazu@umich.edu
	}	
}

\begin{document}
\maketitle    
\begin{abstract}
{\it
This paper presents a multicomponent topology optimization method for designing structures assembled from additively-manufactured components, considering anisotropic material behavior for each component due to its build orientation, distinct material behavior and stress constraint at component interfaces (\textit{i.e.}, joints). Based upon the multicomponent topology optimization (MTO) framework, the simultaneous optimization of structural topology, its partitioning, and the build orientations of each component is achieved, which maximizes an assembly-level structural stiffness performance subject to maximum stress constraints at component interfaces. The build orientations of each component are modeled by its orientation tensor that avoids numerical instability experienced by the conventional angular representation. A new joint model is introduced at component interfaces, which enables the identification of the interface location, the specification of a distinct material tensor, and imposing maximum stress constraints during optimization. Both 2D and 3D numerical examples are presented to illustrate the effect of the build orientation anisotropy and the component interface behavior on the resulting multicomponent assemblies.
}
\end{abstract}

\section{Introduction}
Due to the nature of additive layer manufacturing, where a component is made by adding layer-upon-layer of materials, the resulting structural behavior of manufactured component is anisotropic depending on the chosen build orientation. This is particularly the case for fiber-reinforced additive processes such as the continuous fiber printing and the binding of stacked long fiber sheets ~\cite{markforged,impossibleobject} , where the in-layer mechanical properties are further enhanced with fiber-based reinforcement.

While such anisotropic structural behavior has been experimentally observed (\textit{e.g.},~\cite{frazier2014metal,carroll2015anisotropic,popovich2017functionally}), limited attempts have been made to consider this factor during the computational design optimization of additively-manufactured structures. Ulu \textit{et al.}~\cite{ulu2015enhancing} considered structural anisotropic behaviors while optimizing the build orientations of additively manufactured mechanical products. Their optimization scheme is based on the design of experiments of printed samples and surrogate modeling. To manage the complexity of the designed experiments, the overall product geometry was given a \textit{priori} and fixed throughout the optimization of the build orientations. In addition to anisotropic structural behaviors, the build orientations have impacts on the need of support structures, surface quality, and fabrication time. Related works considering these factors also have primarily used sampling-based non-gradient optimization schemes to determine the optimal build orientations (\textit{e.g.},~\cite{alexander1998part, pandey2004optimal, gao2015revomaker, langelaar2018combined, jaiswal2018build, wu2018general, song2016cofifab}). However, these work can only explore small ranges of possible structures and orientations, since the applicability of sampling-based approaches are intrinsically limited by the available samples. While approaches based on analytic models do not pose this limitation, they often suffer from numerical instability during optimization, especially in 3D, due to the cyclic behavior of trigonometric functions that appear in the models with respect to the build orientation angles.

Topology optimization, as a computational method for exploring the optimal shape of geometries, has become an effective tool for designing components made by additive manufacturing (\textit{e.g.},~\cite{dede2015topology,zegard2016bridging,zhu2017two,vogiatzis2017open}). Conventional topology optimization methods often assume that the optimized structure will be produced as a single piece. This assumption, however, severely limited the design exploration space. Focusing on powder bed additive manufacturing, Zhou \textit{et al.}~\cite{zhou2019jcise} demonstrated that by allowing multiple components in topology optimization, the resulting assembly design is no longer limited by the maximum build volume of a chosen printer and the cavity-free requirements intrinsic to additive manufacturing processes. While~\cite{zhou2019jcise} did not consider the build orientation of each component as design variables, the allowance of multiple build orientations (and the corresponding material anisotropicity) optimized for each component is expected to open further opportunities for assembly-level performance improvement. A challenge is, to overcome the aforementioned numerical difficulty associated with the angular representation of the build orientations, especially for use with gradient-based optimization algorithms.

An inevitable challenge arose from allowing multiple components in topology optimization is the modeling of the distinct component interface (\textit{i.e.}, joint) behavior. Both stiffness-based and strength-based joint models have been investigated in multicomponent topology optimization, with some notable limitations. Early works are limited to discrete representations of component interfaces, which require the use of non-gradient optimization methods~\cite{lyu2005topology,yildiz2011topology,guirguis2015multi}. Prior attempts for interface modeling in a continuous optimization framework have been limited to stiffness-based behavior~\cite{zhou2018mtos,zhou2019jcise,zhou2019mtod}. Relatedly, some attempts have been made recently in the field of multimaterial topology optimization for modeling the interface behavior between different materials~\cite{liu2016multi,woischwill2018multimaterial,liu2018integrated,chu2019topology}, which inspired the new joint model presented in this paper.

Based upon the multicomponent topology optimization (MTO) framework~\cite{zhou2018mtos,zhou2018mtoc,zhou2019jcise,zhou2019mtod}, the methods presented in this paper contribute primarily in two aspects. First, the build orientations of each additively-manufactured component are optimized based on an orientation tensor representation~\cite{nomura2019inverse}, which effectively avoids the numerical instability associated with the angular representation of the build orientations. A transversely isotropic material tensor is used to model the anisotropic material behaviors, where Young's modules in the build direction is lower than the ones in other directions. Second, a new component interface (\textit{i.e.}, joint) model is introduced, which enables the identification of the interface location, the specification of a distinct material tensor, and imposing maximum stress constraints during optimization. Both 2D and 3D numerical examples are presented to illustrate the effect of the build orientation anisotropy and the component interface behavior on the resulting multicomponent assemblies.

\section{Mathematical Model}
\subsection{Design Variables}
Following~\cite{zhou2018mtoc,zhou2019jcise}, density and component membership field design variables are used to design the base topology and partitioning of MTO, respectively. In this paper, additional orientation design variables are introduced to design component-wise build orientations.

\begin{figure}
	\centering
	\includegraphics[width=0.6\textwidth]{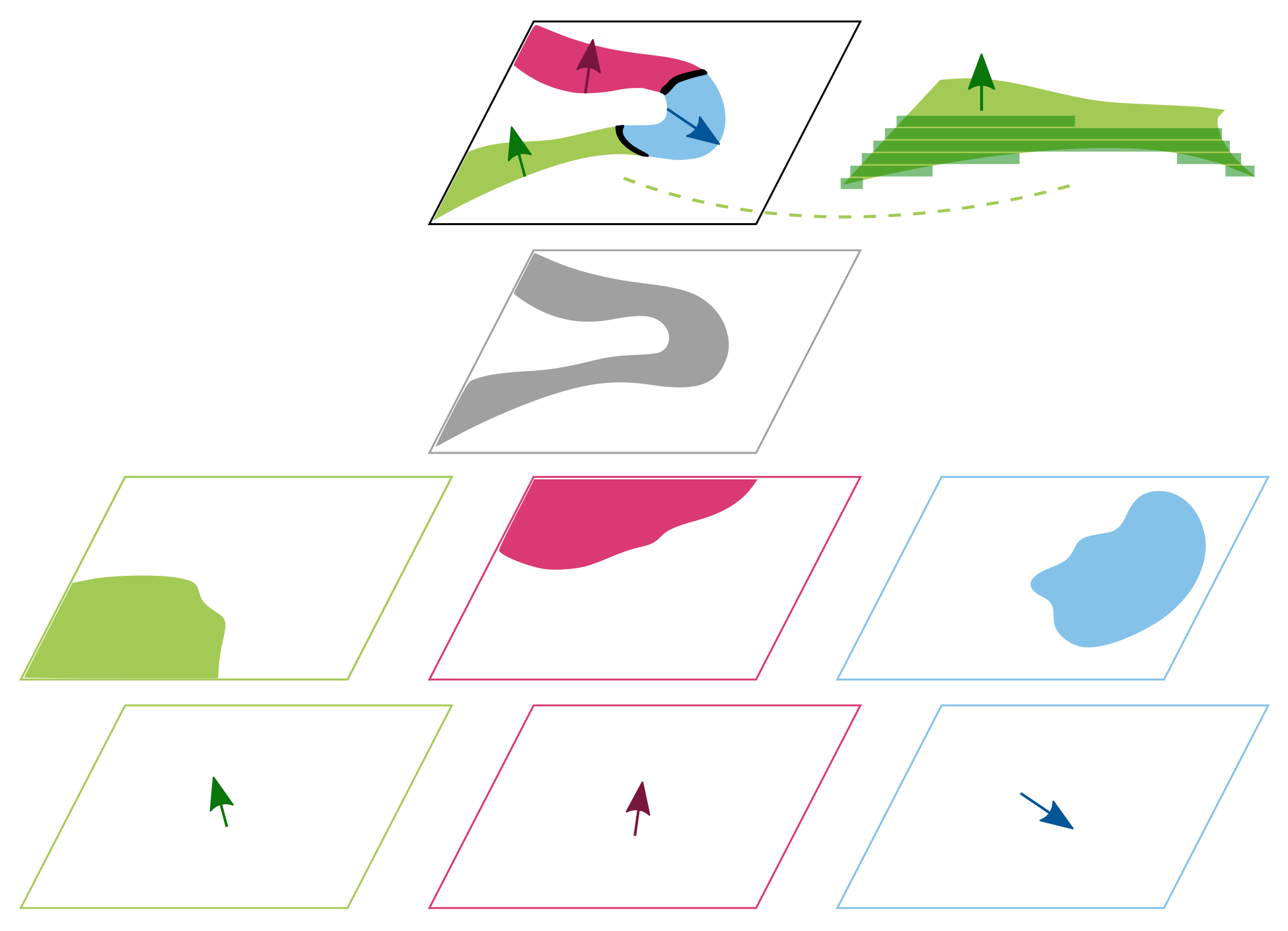}
	\caption{Design variables. From top to bottom: an example multicomponent topology with component-wise build orientations (black arrows) and joints (black lines); density field $\rho$; component membership field $m^{(k)}$; component-wise build orientations $\mathbf{q}^{(k)}$.}
	\label{fig1}
\end{figure}

As seen in Fig.~\ref{fig1}, the density field variable $\rho$ describes the overall base topology, where $1$ indicates solid material and $0$ indicates void. While $\rho$ is a continuous variable ranging between 0 and 1 during the course of optimization, the convergence of 1 or 0 densities (\textit{i.e.}, black or white) at the end of optimization is desired. This is achieved by the classic Solid Isotropic Material with Penalization (SIMP) approach~\cite{bendsoe2004topology}. The component membership vector field design variable $\mathbf{m}=(m^{(1)}, m^{(2)},\cdots,m^{(K)})$ governs the component partitioning, where $m^{(k)}\in[0,1]$ is the fractional membership of a design point to the $k$-th component, $k=1,2,\cdots,K$. Dimension $K$ is the prescribed maximum allowable number of components. While $m^{(k)}$ is a continuous variable ranging between 0 and 1, the clear component partitioning at the end of optimization requires the membership to only one component being 1 and the rest being 0 at convergence. It can be achieved by either the hypercube-to-simplex projection~\cite{zhou2018mtoc} or the Discrete Material Optimization (DMO) projection~\cite{stegmann2005discrete}. The latter is used in this paper. Both $\rho$ and $m^{(k)}$ are regularized following the framework discussed in~\cite{kawamoto2011heaviside}, which includes a Helmholtz partial differential equation filter and a smoothed Heaviside projection method. For the details of implementing the DMO projection for MTO and the design variable regularization, interested readers are referred to~\cite{zhou2019jcise}. The build orientation design is parameterized with the orientation tensor representation~\cite{nomura2019inverse}. The vector design variable $\mathbf{q}^{(k)}$ describes the build orientation of component $k$, as described in Section~\ref{sec:orientationtensor}.

\subsection{Joint Model and Stress Constraint}
Structural assemblies comprised of multiple components require a joining procedure after all components are manufactured separately. Depending on a chosen joining process (\textit{e.g.}, welding, adhesive bonding, and screwing), component interfaces where joints exist often exhibit a different, often inferior, mechanical characteristic from the joined components. For this reason, joint locations are often regarded as critical regions where failures occur. To integrate such interface characteristics into MTO, this section proposes a new joint model that can identify component interfaces, assign distinct joint tensors, and impose maximum stress constraints, during the optimization iterations.

\begin{figure}
	\centering
    \includegraphics[width=0.6\textwidth]{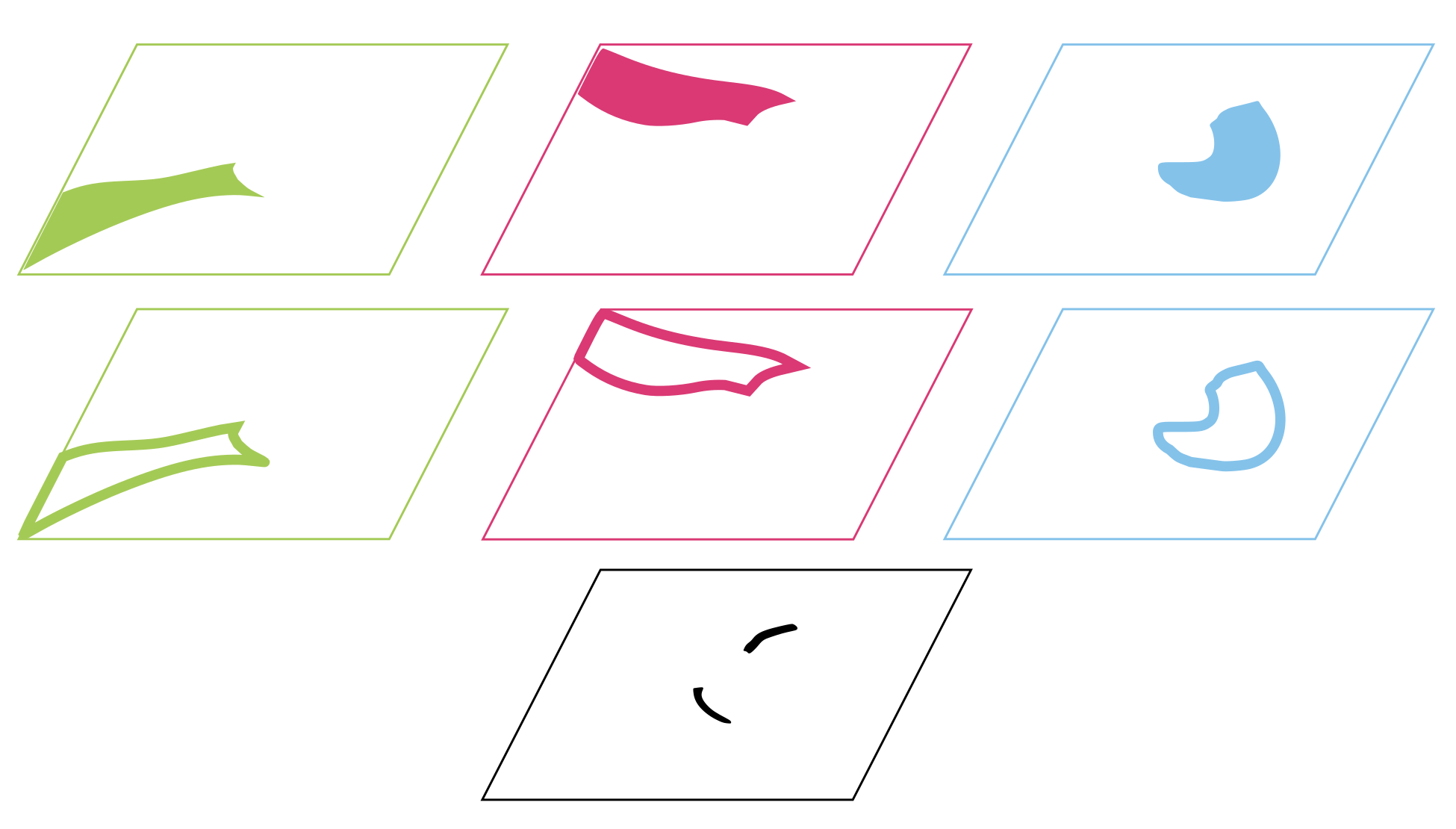}
	\caption{Interface identification. From top to bottom: component described by $\rho^p m^{(k)}$; component boundary identified by $||\nabla(\rho^p m^{(k)})||^2$; interface indicator $I$.}
	\label{fig2}
\end{figure}

In MTO, both the overall base topology (represented by density field $\rho$) and component partitioning (represented by component membership field $\mathbf{m}$) evolve during the course of optimization. The identification of component interfaces is especially challenging when the density and component membership design fields are still ``blurry", which is typically the case until very close to the end of the optimization. Based on the state of density field $\rho$ and component membership field $\mathbf{m}$, interface indicator $I$ is defined at each design point as follows:
\begin{equation}
    I = \hat{H} \left( \sum_{\substack{i,j=1,2,...,K \\ i>j}}  \left\Vert \nabla(\rho^p m^{(i)}) \right\Vert^2 \left\Vert\nabla(\rho^p m^{(j)})\right\Vert^2 \right),
    \label{eq:ji}
\end{equation}

where $\hat{H}$ is a shifted (to a positive side) smoothed Heaviside step function, $\nabla$ is the spatial gradient operator, and $p$ is the SIMP power law penalization parameter. Since expression $||\nabla(\rho^p m^{(k)})||^2$ gives positive values along the boundary of component $k$, the product of $||\nabla(\rho^p m^{(i)})||^2$ and $||\nabla(\rho^p m^{(j)})||^2$ gives positive values along the interface between components $i$ and $j$. The sum of the product for all possible component pairs, therefore, gives positive values along the interfaces between any component pairs. After the Heaviside projection $\hat{H}$, the resulting interface indicator $I$ at a design point equals to $1$ if the point is at a component interface and $0$ otherwise. Figure~\ref{fig2} illustrates the proposed interface identification scheme.

In order to model the distinct mechanical property at component interfaces different from that of the base component materials, user-specified joint stiffness tensor $\mathbf{C}_J$ can be defined based on the chosen joining process. The combined stiffness tensor can be defined at each design point as follows:
\begin{equation}
    \mathbf{C} = (1-I)\mathbf{C}_B \left( \rho,\mathbf{m} \right) + I\mathbf{C}_J,
\end{equation}
where $I$ is the interface indicator defined in Eq.~(\ref{eq:ji}), $\mathbf{C}$ is the combined stiffness tensor, $\mathbf{C}_J$ is the user-specified joint tensor, and $\mathbf{C}_B$ is the transversely isotropic base material tensor. The resulting $\mathbf{C}$ at a design point equals to $\mathbf{C}_B$ when $I = 0$ (not in a component interface) while $\mathbf{C}$ equals to $\mathbf{C}_J$ when $I = 1$ (in a component interface). Due to the smoothed Heaviside operator $\hat{H}$, locations near component interfaces will inevitably have certain mixed properties of $\mathbf{C}_J$ and $\mathbf{C}_B$. While it is acknowledged that this effect cannot be completely eliminated, it can be made negligibly small by continuation scheme to Heaviside parameters.

A maximum stress constraint can then be applied to the identified component interfaces. The implementation of the stress constraint follows the classic P-Q relaxation concept~\cite{le2010stress} and the Heaviside constraint aggregation approach~\cite{wang2018heaviside}: 
\begin{equation}
    \ds\int_{D} I\cdot H \left( \frac{\sigma - \bar{\sigma}}{\bar{\sigma}} \right) \left( \frac{\sigma}{\bar{\sigma}} \right)^2 \dd{\Omega} \leq \bar{\epsilon},
\label{eq:sig}
\end{equation}
where $I$ is the interface indicator defined in Eq.~(\ref{eq:ji}), $H$ is a smoothed Heaviside function, $\sigma$ is the Mises stress, $\bar{\sigma}$ is the user-specified maximum allowable stress threshold, and $\bar{\epsilon}$ is an infinitesimal numerical tuning parameter. Multiplying $I$ guarantees that the maximum stress constraint is applied only to component interfaces, and the smoothed Heaviside function $H$ gives $1$ when $\sigma>\bar{\sigma}$. Therefore Eq.~(\ref{eq:sig}) is satisfied if all stress values in the identified component interfaces are no more than $\bar{\sigma}$.

\subsection{Build Orientation Design}
\label{sec:orientationtensor}
To model the anisotropic mechanical behavior of an additively manufactured component, the transversely isotropic tensor is used. The elements of an orientation tensor are regarded as design variables, subject to the constraints to guarantee the transverse isotropicity. The method described below is based on the orientation tensor method for general material orientation design~\cite{nomura2019inverse}, which effectively avoids the numerical difficulty associated with the angular or vector representation of material orientation, especially in 3D. Interested readers should refer to~\cite{nomura2019inverse} for more details.

Let ${C}_t$ be a transversely isotropic tensor in the 1-2 directions, which can be given in a generic form as:
\begin{equation}
    \mathbf{C}_t = 
    \left({\begin{array}{cccccc}
    C_{1111} & C_{1122} & C_{1122} & 0 & 0 & 0\\
    & C_{2222} & C_{2233} & 0 & 0 & 0\\
    & & C_{2222} & 0 & 0 & 0\\
    & & & C_{2323} & 0 & 0\\
    & \text{sym.} & & & C_{1212} & 0\\
    & & & & & C_{1212}
    \end{array} } \right),
    \label{eq:transverse}
\end{equation}
and let $\mathbf{p}^{(k)}$ be a unit vector, defined for each component $k$. Then, the transversely isotropic tensor ${\mathbf{C}_{r}}^{(k)}=({C_{r}^{(k)}}_{ijkl})$ obtained by rotating $\mathbf{C}_t$ to the orientation defined by $\mathbf{p}^{(k)}$ is given as~\cite{advani1987use}:
\begin{equation}
    \begin{aligned}
    {C_r^{(k)}}_{ijkl}= 
    &B_1 a_{ij}^{(k)}a_{kl}^{(k)} +B_2 (a_{ij}^{(k)}\delta_{kl} + a_{kl}^{(k)}\delta_{ij})+\\
    &B_3(a_{ik}^{(k)}\delta_{jl} + a_{il}^{(k)}\delta_{jk} + a_{jk}^{(k)}\delta_{il} + a_{jl}^{(k)}\delta_{ik})+\\
    &B_4(\delta_{ij}\delta_{kl}) + B_5(\delta_{ik}\delta_{jl}+\delta_{il}\delta_{jk}),
    \end{aligned}
    \label{eq:tensor_rot}
\end{equation}
where $i,j,k,l=1,2,3$. $\delta$ is Kronecker's delta and $(a_{ij}^{(k)})$ is the second order orientation tensor with respect to $\mathbf{p}^{(k)}$:
\begin{equation}
    (a_{ij}^{(k)}) = \mathbf{p}^{(k)}\dyadic\mathbf{p}^{(k)}
    = \left(\begin{array}{ccc}
    a_{11}^{(k)}&a_{12}^{(k)}&a_{13}^{(k)}\\
    &a_{22}^{(k)}&a_{23}^{(k)}\\
    \text{sym.}&&a_{33}^{(k)}\\
    \end{array}\right)
    \label{eq:tensor2}
\end{equation}
and $B_1, B_2, \cdots, B_5$ are coefficients given as:
\begin{subequations}
    \begin{align}
    B_1& =  C_{1111}+C_{2222}-2C_{1122}-4C_{1212}\\
    B_2& =  C_{1122} -C_{2233}\\
    B_3& =  C_{1212}+(C_{2233}-C_{2222})/2\\
    B_4 & =   C_{2233} \\
    B_5 &  =  (C_{2222}-C_{2233})/2.
    \end{align}
\end{subequations}

For each design point, the elements of orientation tensor $a_{ij}^{(k)}$, $i,j=1,2,3$ in Eq.~(\ref{eq:tensor2}) rather than orientation vector $\mathbf{p}^{(k)}$, are used as the build orientation design variables for each component $k$ during optimization. These six variables are not independent and subject to the tensor invariant conditions: 
\begin{subequations}
    \begin{align}
    &a_{11}^{(k)}+ a_{22}^{(k)} + a_{33}^{(k)} = 1\\
    &M_{ii}^{(k)} = 0 \text{\qquad for \quad } i=1,2,3,
    \end{align}
    \label{eq:conditions}
\end{subequations}
where $M_{ii}^{(k)}$ is the second order principal minor determinants of $(a_{ij}^{(k)})$.

Instead of directly varying $a_{ij}^{(k)}$ subject to the constraints in Eq.~(\ref{eq:conditions}), the optimizer will use another vector $\mathbf{q}^{(k)}$ of six unconstrained variables defined for each design point and each component $k$, which are mapped to $a_{ij}^{(k)}$:
\begin{equation}
    \mathbf{q}^{(k)}=(q_{11}^{(k)},q_{22}^{(k)},q_{33}^{(k)},q_{12}^{(k)},q_{13}^{(k)},q_{23}^{(k)})
\end{equation}
where $q_{ij}^{(k)}\in[\delta_{ij}-1,1]$.  To satisfy the first tensor invariant condition in Eq.~(\ref{eq:conditions}a), the first three variables $q_{11}^{(k)}$, $q_{22}^{(k)}$, and $q_{33}^{(k)}$ are mapped to $a_{11}^{(k)}$, $a_{22}^{(k)}$, and $a_{33}^{(k)}$ through the cube-to-simplex (\textit{i.e.}, hexahedral-to-tetrahedral) projection method~\cite{zhou2018mtoc,nomura2019inverse}. To satisfy the second tensor invariant condition in Eq.~(\ref{eq:conditions}b), the second three variables $ q_{12}^{(k)}$, $q_{13}^{(k)}$, and $q_{23}^{(k)}$ are mapped to $a_{12}^{(k)}$, $a_{13}^{(k)}$, and $a_{23}^{(k)}$ through a Heaviside projection method:
\begin{equation}
    a_{ij}^{(k)} = \left\{2\tilde{H}( q_{ij}^{(k)})- 1 \right\} \sqrt{a_{ii}^{(k)}a_{jj}^{(k)}},
\end{equation}
where $\tilde{H}$ is a smoothed Heaviside function. 

Finally, the combined base material tensor for each design point is given as a function of $\mathbf{q}^{(k)}$, as follows:
\begin{equation}
    \mathbf{C}_B(\mathbf{q}^{(k)}) = \rho^P \sum_{k=1}^K m^{(k)}\mathbf{C}_r^{(k)}(\mathbf{q}^{(k)}),
\end{equation}
where $\mathbf{C}_{r}^{(k)}$ is the rotated transversely isotropic tensor for component $k$ in Eq.~(\ref{eq:tensor_rot}).

\subsection{Optimization Formulation}
The overall anisotropic MTO considering component-wise build orientations, subject to a volume fraction constraint and a maximum stress constraint at component interfaces can be summarized as follows:
\begin{equation}
	\begin{aligned}
	&\underset{ \substack{\rho,\mathbf{m} \\ \mathbf{q}^{(1)},\cdots,\mathbf{q}^{(K)}}}{\text{minimize}} && F := \int_D\frac{1}{2}\boldsymbol{\sigma}^{\intercal} \boldsymbol{\epsilon}\dd\Omega \\
	&\text{\text{subject to}} && g_1 := \int_{D}\rho \dd{\Omega}/V_0 - \bar{V} \leq 0 \\
	&{} && g_2 := \ds\int_{D} I\cdot H \left( \frac{\sigma - \bar{\sigma}}{\bar{\sigma}} \right) \left( \frac{\sigma}{\bar{\sigma}} \right)^2 \dd{\Omega} - \bar{\epsilon} \leq 0 \\
	& {} && \rho\in[0,1]^D \\
	&{} && \mathbf{m}\in[0,1]^{K\times D}\\
	\\
	&{}	&&	\text{for } k=1,2,\cdots,K \text{ and } \\ 
	&{}	&&	(i,j)\in\{(1,1),(2,2),(3,3),(1,2),(1,3),(2,3)\}:\\
	&{} && q_{ij}^{(k)}\in[\delta_{ij}-1,1]
	\end{aligned},
	\label{eq:optimization}
\end{equation}
where $\rho$ and $\mathbf{m}$ are the density and component membership field design variables, respectively, $\mathbf{q}^{(k)}$ is the build orientation design variables for component $k$, and $K$ is the prescribed maximum allowable number of build orientations. Objective function $F$ is the structural compliance of the overall assembly. Constraint $g_1$ is the volume fraction constraint, where $V_0$ is the total volume of design domain $D$ and $\bar{V}$ is the volume fraction constraint limit. Constraint $g_2$ is the maximum stress constraint on component interfaces, where $\bar{\sigma}$ is the maximum allowable stress and $\bar{\epsilon}$ is an infinitesimal constraint limit. The stress field $\boldsymbol{\sigma}$ and the strain field $\boldsymbol{\epsilon}$ are obtained by solving linear elasticity static equilibrium equations.

The optimized number of build orientations can converge to an integer equal to or less than the prescribed $K$. It is noted that the resulting number of components can be greater than $K$ because multiple disconnected components can be generated for each build orientation phase $k$. While the formulation in Eq.~\ref{eq:optimization} assumes there is only one build orientation for each component, this can be readily generalized to the situation where one component can be built with multiple build orientations, as in the case of certain advanced printing equipment with variable-orientation platforms.

The formulation relies on the anisotropic material property caused by build orientations as a major driver for component partitioning, rather than process-specific geometric constraints as done in some of our past MTO work~\cite{zhou2018mtos,zhou2019jcise,zhou2019mtod}. In particular, it is decided to not to include the constraints on printer size~\cite{zhou2019jcise} in this initial attempt, in order to manage the increased numerical challenges associated with the addition of the build orientation variables. The inclusion of printer size constraints will be one of the immediate future work.

\section{Numerical Examples}
\begin{figure}
	\centering
    \includegraphics{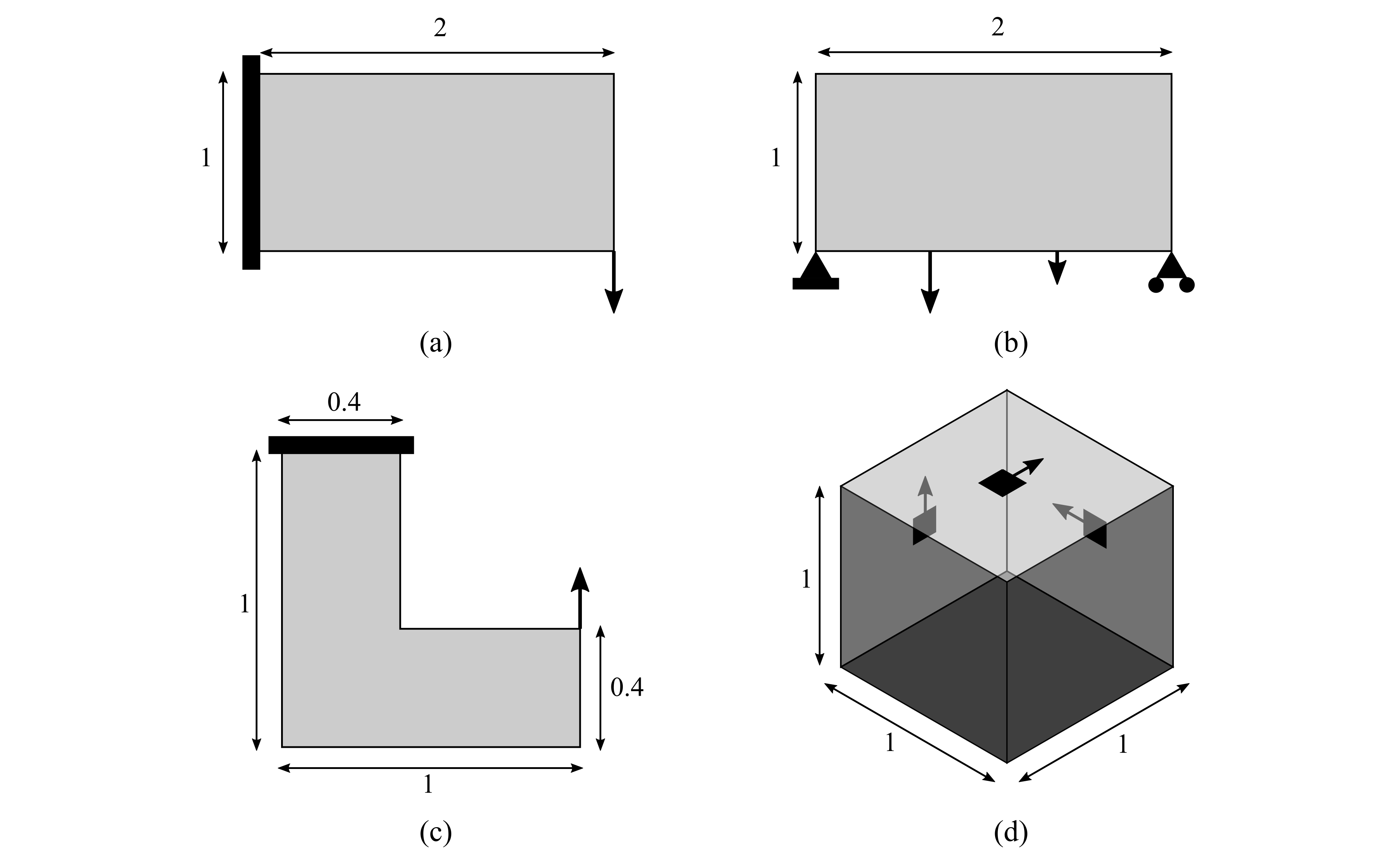}
	\caption{Design domains and boundary conditions for the numerical examples of (a) 2D cantilever; (b) 2D bridge; (c) 2D L-bracket; and (d) 3D multiple loading.}
	\label{fig3}
\end{figure}

This section presents 2D and 3D numerical examples. Their design domain and boundary condition settings are summarized in Fig.~\ref{fig3}. Figure~\ref{fig3}~(a) is a 2D cantilever example, which is used to compare the anisotropic multicomponent design with isotropic single-piece design. Figure~\ref{fig3}~(b) is a 2D bridge example, which is used to demonstrate the effect of joint stiffness specification on multicomponent designs. Figure~\ref{fig3}~(c) is an 2D L-bracket example, which is used to demonstrate the effect of component interface stress constraint on multicomponent designs. Finally, Fig.~\ref{fig3}(d) is a 3D multiple loading example, showcasing the applicability of the proposed method to more complicated problems.

The continuous nonlinear constrained optimization problem in Eq.~(\ref{eq:optimization}) was solved by the method of moving asymptotes (MMA)~\cite{svanberg1987method}. The sensitivity analysis followed a standard adjoint method and was implemented using COMSOL Multiphysics. The optimizer used only the first-order gradient. The density and component membership design fields were uniformly initialized. The build orientation design variables were randomly initialized using a multi-start initialization scheme to avoid inferior local solutions. The optimizer terminates when either the change of the objective function value stays within the prescribed lower bound or the maximum number of iterations is reached. The volume fraction upper bound $\bar{V}$ for all examples was set to $0.35$.

The transversely isotropic tensor requires five independent constants, namely $\left( E_1, E_2, \nu_{21}, \nu_{23}, G_{12}\right)$. Notably, $E_1$ determines the Young's modulus along the build direction. $E_2$ determines the Young's modulus for other directions. For additive layer manufacturing, $E_1$ is smaller than $E_2$. For the sake of simplicity, the joint tensor $\mathbf{C}_J$ is assumed isotropic in all numerical examples, which is governed by its Young's modulus $E_j$ and Poisson ratio $\nu_j$. Anisotropic joint model differentiating shear and tensile directions is left for future research.

Table~\ref{tab:para} summarizes shared parameters for all numerical examples, unless otherwise noted. All material properties, loads and dimensions are given in relative (unitless) measures.
\begin{table}[t]
	\caption{Summary of parameters}
	\begin{center}
	\label{tab:para}
	\begin{tabular}{c l l}
		\hline
		Symbol & Value & Description \\
		\hline
		$E_1$			    &$2$    &Young's modulus along build direction\\
		$E_2$               &$10$   &Young's modulus in other directions\\
		$\nu_{21}$		    &$0.3$  &Poisson ratio\\
		$\nu_{23}$		    &$0.3$  &Poisson ratio\\
		$G_{12}$		    &$3.85$ &shear modulus\\
		$E_j$               &$1$    &Young's modulus of joint (except Example 2)\\
		$\nu_j$             &$0.3$  &Poisson ratio of joint\\
		$P$		            &$3$    &SIMP penalization\\
		$\bar{V}$           &$0.35$ &volume fraction constraint limit\\
		$\bar{\epsilon}$    &$0.01$ &infinitesimal constraint limit\\
		\hline
	\end{tabular}
	\end{center}
\end{table}

\subsection{2D Cantilever}
To compare the anisotropic multicomponent design with the conventional isotropic single-piece design, a 2D cantilever example is used. Its design domain and boundary condition settings are presented in Fig.~\ref{fig3}~(a). The left edge is fixed in all degrees of freedom. In this example, the maximum allowable stress at component interfaces $\bar{\sigma}$ is set to a large value $1000$ (equivalent to unbounded). The number of allowable build orientations $K$ is set to $3$.

\begin{figure}
	\centering
    \includegraphics{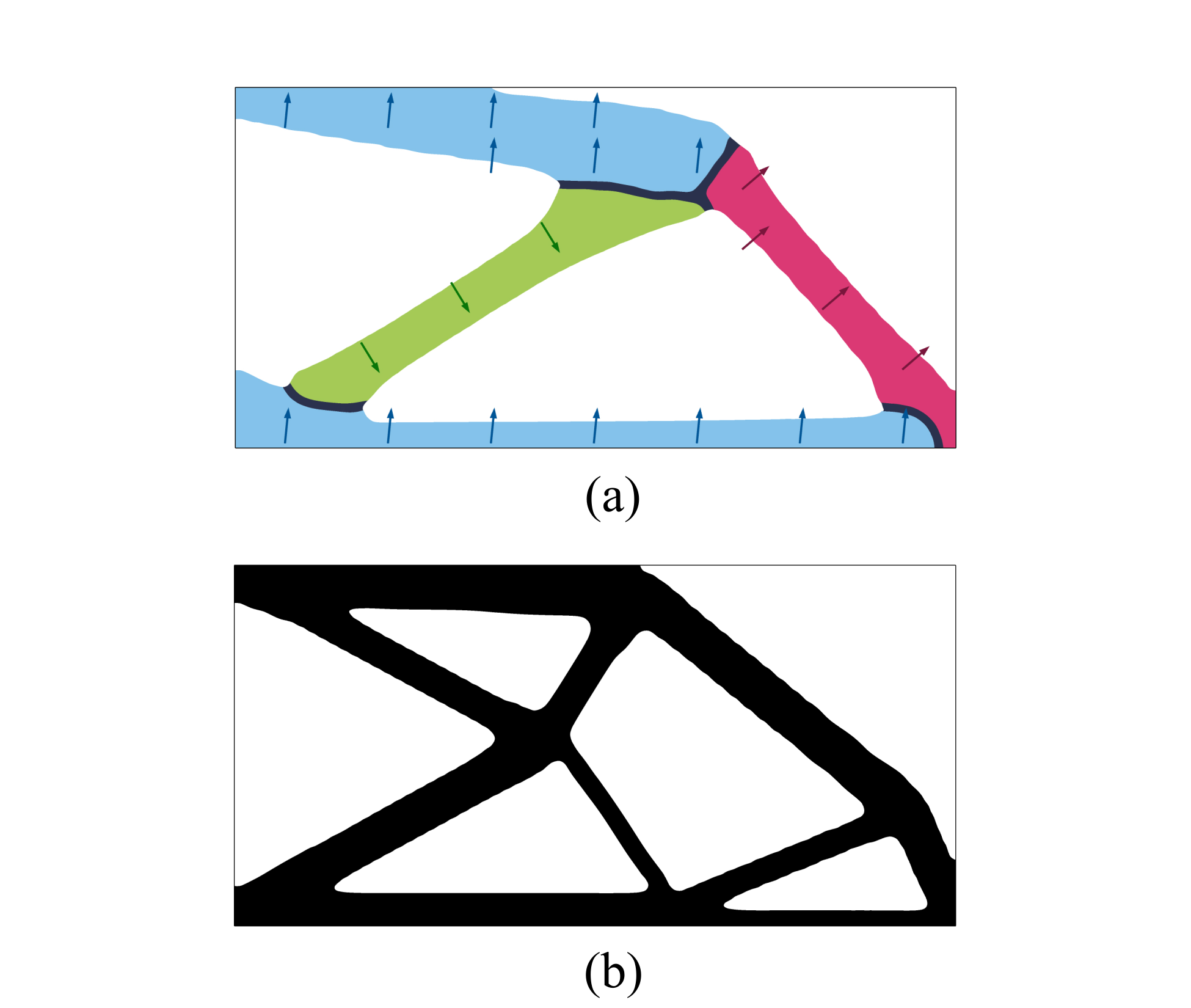}
	\caption{(a) The optimized anisotropic multicomponent cantilever design with build orientations and joints. Arrows indicate the optimized build orientations. (b) The optimized isotropic single-piece design.}
	\label{fig4}
\end{figure}

The optimized multicomponent cantilever design is presented in Fig.~\ref{fig4}~(a). With three distinct build orientations, the resulting multicomponent assembly has four individual components. The arrows on components indicate the optimized build orientations. Each component is also colored according to its build orientation. The blank lines between the colored components indicate the thin layers of component interfaces (joint) regions. It is noted the resulting number of components can be greater than the prescribed number of build orientations, since multiple separated components can possibly share a single build orientation, like the two blue components in~\ref{fig4}~(a)). The optimized component-wise build orientations are mostly perpendicular to the major principal stress directions, which supports the empirical mechanics knowledge for maximizing the assembly-level structural stiffness. The components can be printed either separately based on their corresponding optimized build orientations or together after proper alignments on the printer base plate. 

Figure~\ref{fig4}~(b) presents the isotropic single-piece design with the same volume fraction setting. The isotropic single-piece design has a different topology, which is comprised of larger number of thinner bars. The compliance objectives of the two designs are not reported because their relative performance depends on the material selection (\textit{i.e.,} Young's modulus) of the isotropic design, which hinders fair comparison of numerical values.

\subsection{2D Bridge: Joint Stiffness}
To examine the effect of joint stiffness on the optimized base topology and component partitioning, a 2D bridge example is used. Its design domain and boundary condition settings are presented in Fig.~\ref{fig3}~(b). The load applied on the left is twice as much as the load applied on the right. The lower left corner is fixed in all degrees of freedom while the lower right corner is only fixed vertically. To isolate the effect of joint stiffness on the optimized structures, the maximum allowable stress at component interfaces $\bar{\sigma}$ is again set to a large value $1000$ (equivalent to unbounded). The number of allowable build orientations $K$ is set to $2$.

\begin{figure}
	\centering
    \includegraphics{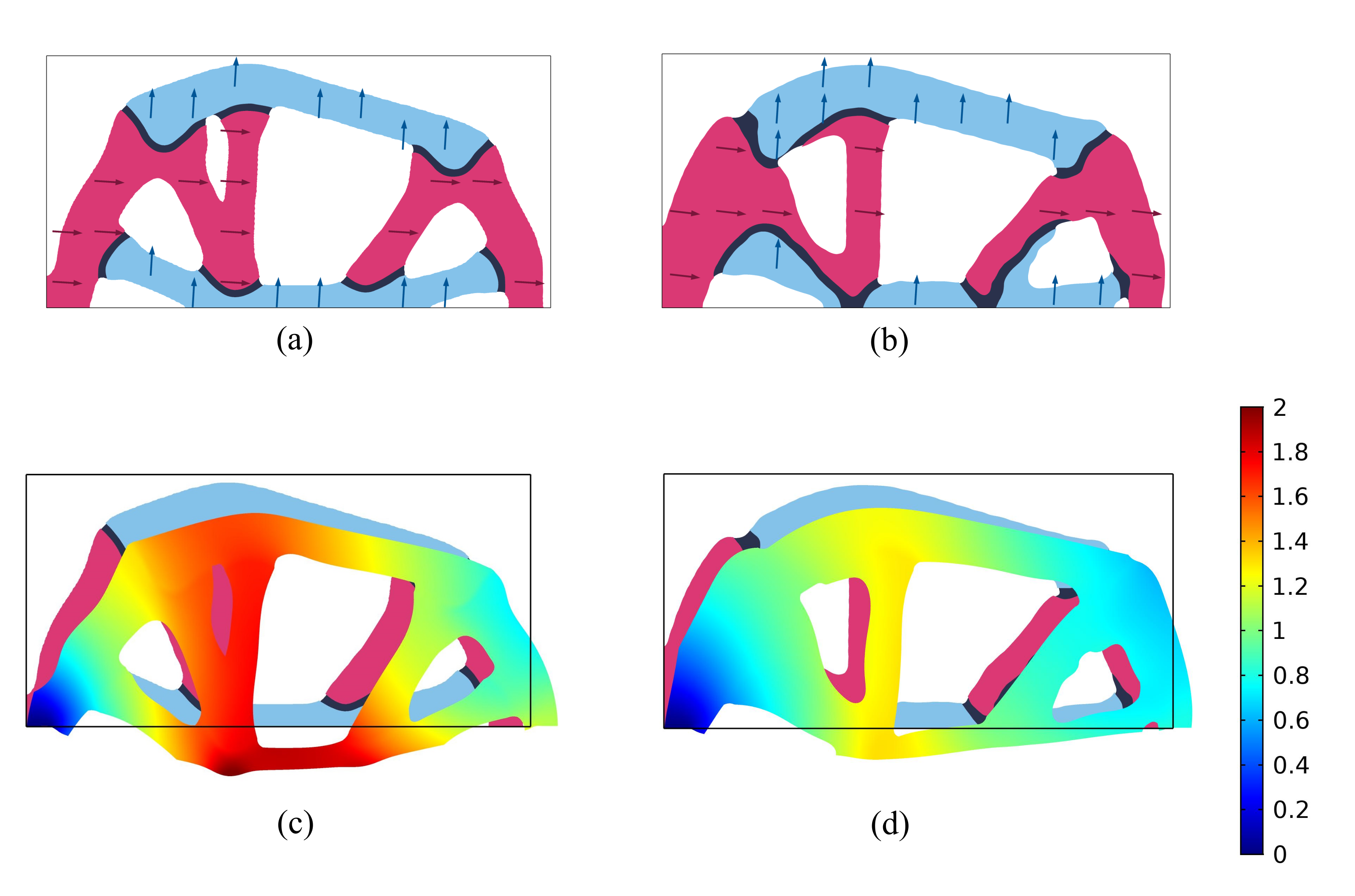}
	\caption{The optimized multicomponent bridge designs with (a) less stiff joints (b) stiffer joints. The displacement field of the deformed bridge designs with (c) less stiff joints (d) stiffer joints. Arrows indicate the optimized build orientations.}
	\label{fig5}
\end{figure}

Two prescribed values of isotropic Young's modulus $E_j$ at joints are used. Figure~\ref{fig5}~(a) presents the optimized multicomponent design with $E_j = 1$, smaller than the Young's modulus of the components in the build direction. Figure~\ref{fig5}~(b) presents the optimized multicomponent design with $E_j = 16$, larger than the Young's modulus of the components in the directions other than the build direction. Two designs have different base topologies and joint configurations. Due to the stiffer joint property, the design in Fig.~\ref{fig5}~(b) allocates the joint material at the loading location. The joint length is also notably longer to take advantage of the stiffer joint property. The resulting assembly-level compliance objectives are $1.36$ and $0.69$, respectively. Due to the stiffer joint used in the design in Fig.~\ref{fig5}~(b), its assembly-level compliance is notably smaller. The displacement field plots of the deformed bridge designs are presented in Figs.~\ref{fig5}~(c) and (d).

\subsection{2D L-Bracket: Joint Stress}
To examine the effect of the maximum stress constraints at component interfaces on the multicomponent assembly design, a 2D L-bracket example is used. Its design domain and boundary condition settings are presented in Fig.~\ref{fig3}~(c). The upper edge is fixed in all degrees of freedom. It is well known that the sharp corner of ``L" shape is where the stress concentration is likely to occur. The maximum allowable stress at component interfaces $\bar{\sigma}$ is set to 1000 (equivalent to unbounded) and 15 for two comparative studies. As all numerical examples use unitless settings, the stress values do not directly relate to practical material properties. The number of allowable build orientations $K$ is set to $3$.

\begin{figure}
	\centering
    \includegraphics{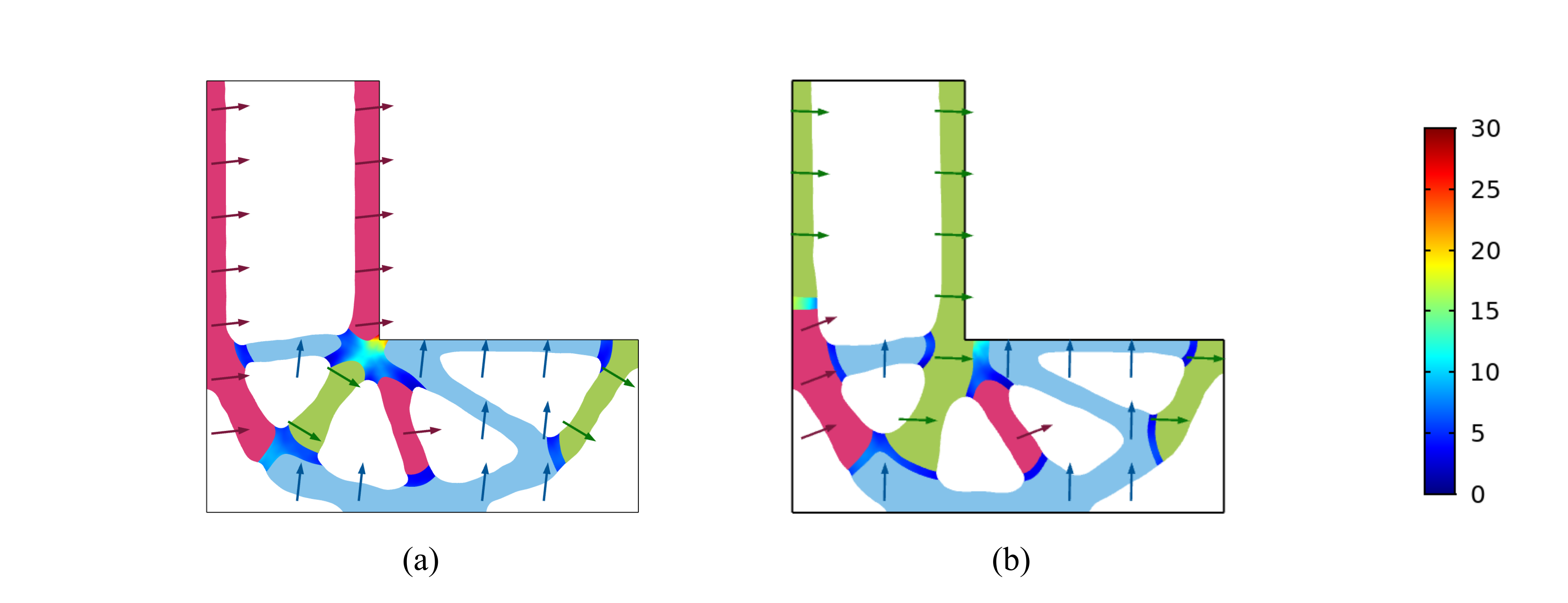}
	\caption{The optimized multicomponent L-bracket designs (a) without the joint maximum stress constraint and (b) with the joint maximum stress constraint. The stress field is plotted at component interfaces. Arrows indicate the optimized build orientations.}
	\label{fig6}
\end{figure}

Figure~\ref{fig6} presents the optimized multicomponent L-bracket designs, with the stress field plotted at component interfaces. For the case of ``unbounded" maximum stress in Fig.~\ref{fig6}~(a), joints appear at a high stress concentration location around the L-bracket corner. On the other hand, with a more strict setting for the maximum allowable stress at component interfaces, Fig.~\ref{fig6}~(b) shows a design that avoids the formation of joints at high stress concentration regions. This example demonstrates the effectiveness of the maximum stress constraint at component interfaces. The assembly design that has a lower maximum stress at component interfaces is expected to be better guarded against joint failures and to function more reliably in real-world environment. The resultant compliance objectives are $5.24$ and $5.36$, respectively. With the more strict allowable maximum stress at component interfaces, the design in Fig.~\ref{fig6}~(b) sacrifices moderately its assembly-level stiffness performance.

\subsection{3D Multiple Load}
To demonstrate the proposed method for more complex 3D structural design problems, a multi-load example is studied. Its cubic design domain and multi-load boundary condition settings are described in Fig.~\ref{fig3}~(d). Three independent shear loads are applied while the opposite faces are fixed in all degrees of freedom. The overall assembly-level objective is defined as the sum of all three independently solved compliance results. The maximum allowable stress at component interfaces $\bar{\sigma}$ is set to a large value $1000$ (equivalent to unbounded). The number of allowable build orientations $K$ is set to $3$.

\begin{figure}
	\centering
    \includegraphics{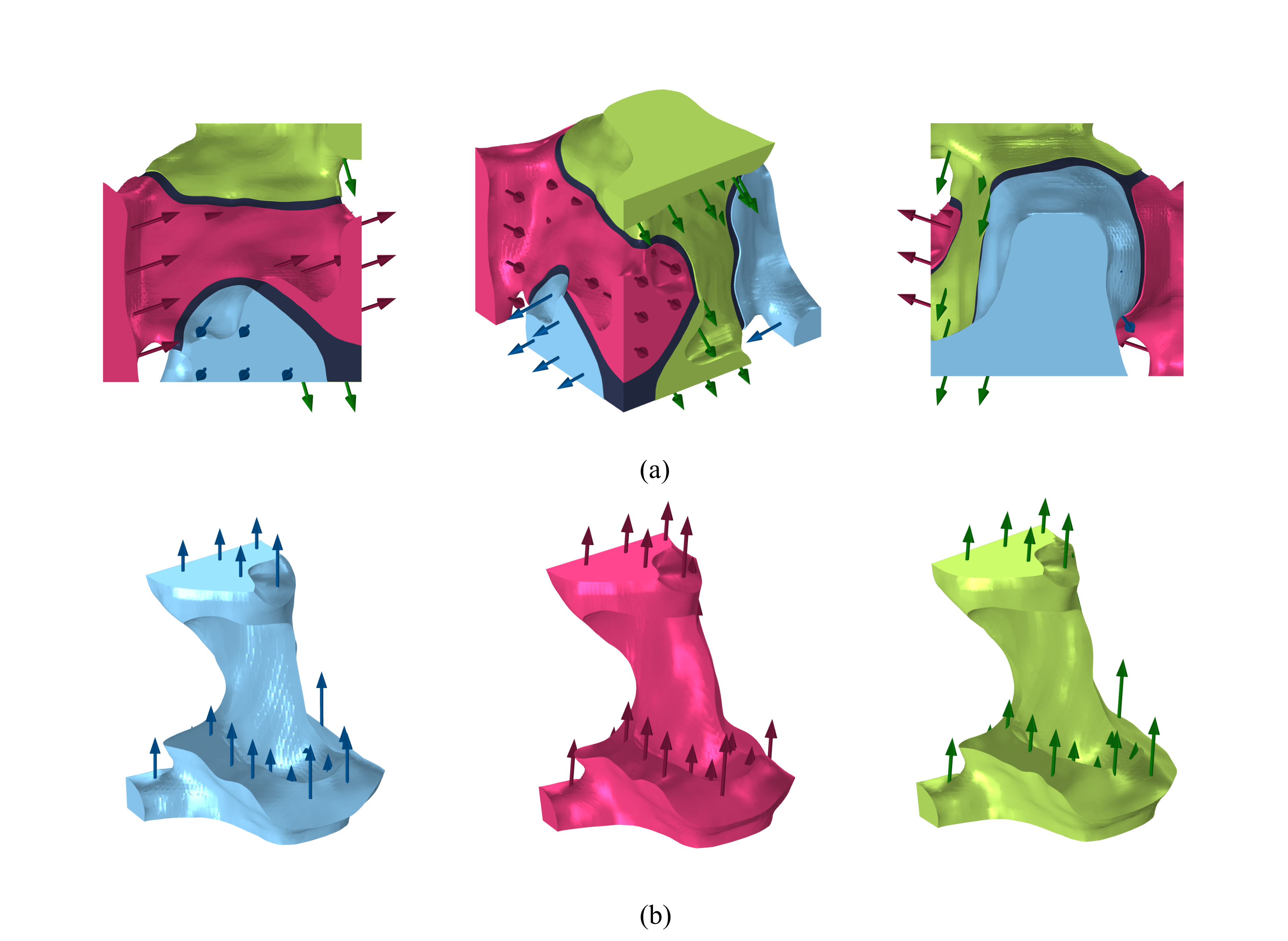}
	\caption{(a) The 3D multicomponent design optimized for a multiple loading condition. (b) The three decomposed components aligned with their corresponding build orientations. Arrows indicate the optimized build orientations.}
	\label{fig7}
\end{figure}

Figure~\ref{fig7}~(a) presents the optimized multicomponent assembly design for the 3D multi-load example. The resulting three-component design has very complex geometry and component interfaces. In particular, it is a hollow structure with a fully enclosing outer shell formed by the three components, which is impossible to additively-manufacture as a single piece. While symmetry is not enforced in the optimization, the three optimized components have almost identical geometries and build orientations due to symmetric boundary conditions, as seen in Fig.~\ref{fig7}~(b).

\section{Conclusion}
This paper presented a multicomponent topology optimization method for additive manufacturing (MTO-A) considering build orientation design and component interface modeling. The proposed method used the anisotropic material property caused by build orientations as a major driver for component partitioning. The optimal design for build orientations was achieved by an orientation tensor optimization method~\cite{nomura2019inverse}. While the stiffness-based joint behavior was considered in~\cite{zhou2019jcise} in a limited fashion, this paper proposed a more comprehensive joint model, which enabled the specification of distinct material properties and a maximum stress constraint applied only at component interfaces. The proposed joint model is generic and can be seamlessly integrated into the other MTO frameworks for different manufacturing processes, such as sheet metal stamping~\cite{zhou2018mtos}, die casting~\cite{zhou2019mtod}, and composite manufacturing~\cite{zhou2018mtoc}. Two 2D examples demonstrated how the build orientation anisotropy and the component interface behavior could affect both the overall base topology and its partitioning. A 3D multicomponent design was optimized for a multi-load condition, which demonstrated the applicability of the proposed method for more complex and practical structural design problems.

Immediate future work would be the integration of constraints on component geometry for additive manufacturing, such as size, enclosed cavity, and overhang. Possible future work includes the direct control of the interface thickness (currently indirectly controlled by filtering and Heaviside parameters), the addition of interface length as a model of assembly cost (currently not considered), and the experimental validation of the optimization results.

\newpage
\listoffigures
\listoftables

\end{document}